\documentclass[12pt]{iopart} % use larger type; default would be 10pt

\usepackage{graphicx}
\usepackage{cite}
\usepackage{hyperref}

\begin{document}
\title{Gravitational Geons on the Brane}
\author{Danny Kermode and Dan Vollick} 
\address{Irving K. Barber School of Arts and Sciences \\
	University of British Columbia Okanagan \\
	3333 University Way, Kelowna, BC V1V 1V7, Canada} 
%\date{} 

\begin{abstract}
In this paper, we examine the possibility of static, spherically symmetric gravitational geons on a 3 dimensional brane embedded in a 4+1 dimensional space-time. We choose a specific \begin{math}g_{tt}\end{math} for the brane-world space-time metric.  We then calculate \begin{math}g_{rr}\end{math} analytically in the weak field limit and numerically for stronger fields.  We show that the induced field equations on the brane do admit gravitational geon solutions.
\end{abstract}


\section{Introduction}
\label{intro}

General Relativity in 3+1 dimensions does not admit static, nonsingular, asymptotically flat and topolotically trivial vacuum space-time solutions \cite{Serini:1918,Einstein:1943,Lichnerowicz:1955}.  In this article, we investigate whether such configurations of the gravitation field, or \emph{gravitational geons} \cite{Wheeler:1955zz,Brill:1964zz,Sones:2005gn}, are possible \emph{on the brane} in the Randall and Sundrum model \cite{Randall:1999vf,Randall:1999ee} in which the three spatial dimensions that we experience are a 3 dimensional brane embedded in a 4+1 dimensional space-time. The effective 3+1 dimensional Einstein field equations on the brane for the Randall and Sundrum model were derived by Shiromizu, Maeda, and Sasaki \cite{Shiromizu:1999wj}.

Gravity on the brane is prevented from \emph{leaking} into the extra dimension at low energies by a negative bulk cosmological constant \cite{Maartens:2010ar}.  This negative cosmological constant can be offset by a positive brane tension such that there is a zero cosmological constant on the brane.  In the case of a gravitational geon solution in which no matter is present, there would nevertheless be a generally non-zero effective energy-momentum tensor induced on the brane corresponding to

\begin{equation}
^{(4)}\mathcal{E}_{\mu\nu}={\frac{1}{8\pi}}{}^{(5)}C^{\alpha}_{\beta\rho\sigma}\eta_{\alpha}\eta^{\rho}h^{\beta}_{\mu}h^{\sigma}_{\nu}
\label{introduction:14}
\end{equation}

\noindent where \begin{math}^{(4)}\mathcal{E}_{\mu\nu}\end{math} is a 4-dimensional projection of the 5-dimensional Weyl tensor \begin{math}^{(5)}C^{\alpha}_{\beta\rho\sigma}\end{math}, \begin{math}\eta_{\alpha}\end{math} is normal to the brane and \begin{math}h_{\mu\nu}=g_{\mu\nu}-\eta_{\mu}\eta_{\nu}\end{math} is the induced metric on the brane \cite{Shiromizu:1999wj,Vollick:2000jv}.  

In their investigation of black holes on the brane, Dadhich, Maartens, Papadopoulos and Rezania showed that the Reissner-Nordstrom metric is an exact solution to the effective 3+1 dimensional Einstein field equations on the brane and can be interpreted as a black hole without electric charge but with a \emph{tidal charge} that arises from the solution to the Einstein field equations in 4+1 dimensions \cite{Dadhich:2000am}.  They further showed that, in the induced metric for their exact black hole solution, the free gravitational field in the bulk contributes a negative effective energy density on the brane and tends to strengthen the gravitational field.

In this paper, we investigate whether the free gravitational field in the bulk modifies the gravitational field on the brane so that gravitational geons are possible on the brane.  Our approach is to look at a specific static, spherically symmetric solution of the form
\begin{equation}
ds^2=-B(r)dt^2+A(r) dr^ 2+ r^2(d\theta^2 + \sin^2\theta d\phi^2) \mbox{    .}
\label{introduction:1}
\end{equation}

\noindent For this metric, the 4 dimensional Ricci scalar on the brane is given by \cite{Vollick:2000jv}
\begin{equation}
R= \frac{B''}{AB} - \frac{B'}{2AB}\left(\frac{A'}{A}+ \frac{B'}{B}\right) + \frac{2}{Ar}\left(\frac{B'}{B} - \frac{A'}{A}\right) + \frac{2}{Ar^2} - \frac{2}{r^2} \mbox{    ,}
\label{introduction:2}
\end{equation}

\noindent where the prime indicates differentiation with respect to r.

Our approach is to choose \begin{math}B(r)\end{math} and solve \begin{math}R=0\end{math} in (\ref{introduction:2}) for \begin{math}A(r)\end{math}.  Substituting a solution to \begin{math}R=0\end{math} into the Einstein field equations for 3+1 dimensional general relativity gives a trace free energy-momentum tensor

\begin{equation}
T_{\mu\nu} = \frac{1}{8\pi} R_{\mu\nu} \mbox{   ,}
\label{introduction:17}
\end{equation}

\noindent that satisfies  

\begin{equation}
\nabla^{\mu}{T}_{\mu\nu} = 0  \mbox{    .} 
\label{introduction:18}
\end{equation}

\noindent For a vacuum solution on the brane, the projection of the 5-dimensional Weyl tensor (\ref{introduction:14}) satisfies \cite{Dadhich:2000am,Maartens:2000fg} 

\begin{equation}
\nabla^{\mu}\mathcal{E}_{\mu\nu} = 0  \mbox{   ,} 
\label{introduction:19}
\end{equation}

\noindent allowing us to make the correspondence \begin{math}8\pi T_{\mu\nu} = -\mathcal{E}_{\mu\nu}\end{math} and treat \begin{math}R=0\end{math} along with equations (\ref{introduction:17}) and (\ref{introduction:18}) as a closed system of equations on the brane.  The effect is that (from \cite{Dadhich:2000am}) \begin{quote}\emph{"a stationary general relativity solution with trace-free energy-momentum tensor gives rise to a vacuum brane-world solution in 5-dimensional gravity."}\end{quote}  Examples of other papers that use this correspondence to investigate vacuum solutions on the brane with \begin{math}R=0\end{math} (or \begin{math}R=4\Lambda\end{math} if one assumes a non-zero cosmological constant on the brane) include \cite{Vollick:2000jv,Bronnikov:2002rn,Casadio:2001jg,Molina:2010yu,Aliev:2009cg,Sheykhi:2008et}.

In this paper we will show that gravitational geons are possible on the brane.

We are not interested in space-time solutions with singularities so we examined the Kretschmann scalar for divergent behaviour in order to identify candidate space-times.  The Kretschmann scalar is given by

\begin{equation}
R_{\mu\nu\rho\sigma}R^{\mu\nu\rho\sigma} = 4K_1^2+8K_2^2+8K_3^2+4K_4^2
\label{introduction:8}
\end{equation}

\noindent where \cite{Bronnikov:2002rn}

\begin{equation}
K_1=\frac{1}{A}\left(\frac{B''}{2B}-\frac{(B')^2}{4B^2}-\frac{A'B'}{4AB}\right) \mbox{    ,}
\label{introduction:9}
\end{equation}

\begin{equation}
K_2=\frac{B'}{2ABr} \mbox{    ,}
\label{introduction:10}
\end{equation}

\begin{equation}
K_3=\frac{-A'}{2A^2r} \mbox{    and}
\label{introduction:11}
\end{equation}

\begin{equation}
K_4=\frac{A-1}{Ar^2} \mbox{    .}
\label{introduction:12}
\end{equation}

From (\ref{introduction:11}) and (\ref{introduction:12}) we must have \begin{math}\displaystyle\lim_{r \to 0}A(r) = 1 + O(r^n) \mbox{ with } n \ge 2\end{math} to avoid divergence of the Kretschmann scalar.  Similarly, from (\ref{introduction:10}) it follows that \begin{math}\displaystyle \lim_{r \to 0}\frac{B'(r)}{B(r)} = O(r^m) \mbox{ with } m \ge 1\end{math}.

In this paper, we take

\begin{equation}
B(r)=1-\frac{2mr^2}{r^3+2ml^2} \mbox{    ,} 
\label{introduction:13}
\end{equation}

\noindent for which \begin{math}B'(0) = 0\end{math}, \begin{math}B(0) = 1\end{math}, both \begin{math}B(r)\end{math} and \begin{math}B'(r)\end{math} are continuous for all \begin{math}r \ge 0\end{math} and \begin{math}B(r)\end{math} is Schwarzschild in the limit as \begin{math}r \to \infty\end{math}.  We also consider general functions \begin{math}B(r)\end{math} in the weak field limit.

The solutions that we investigate have \begin{math}A(r)\end{math} and \begin{math}B(r)\end{math} greater than zero and finite for all \begin{math}r \ge 0\end{math}. This ensures that the spacetime will be nonsingular and without horizons.

\section{Outline of Our Approach}

For static solutions, we have \begin{math}R=0\end{math} which can be written as

\begin{equation}
A'= \frac{(2BB''r^2 - (B')^2r^2 + 4BB'r + 4B^2)A - 4B^2A^2}{Br(B'r + 4B)} \mbox{    .}
\label{approach:1}
\end{equation}

We did not find a general solution to (\ref{approach:1}) for our choice of \begin{math}B(r)\end{math}.  In the absence of a general solution, our approach is as follows

\begin{itemize}
\item
Solve (\ref{approach:1}) in the weak field limit.
\item
Use directional field plots (generated using Maple) and iterative numerical analysis to discover the behaviour of \begin{math}A(r)\end{math}.
\item
Find solutions near potential singular points (i.e. at \begin{math}r=0 \mbox{ and } B'r+4B=0\end{math}) in (\ref{approach:1}) using series approximation.  Specifically, we write (\ref{approach:1}) in the form

\begin{equation}
A'(r)= f(r) A(r) + g(r) A^2(r) \mbox{    .}
\label{approach:2}
\end{equation}

Letting \begin{math}F'(r) = f(r)\end{math} then (\ref{approach:2}) has the solution \cite{Bronnikov:2002rn}

\begin{equation}
A(r)= \frac{-1}{e^{-F(r)}\int{g(r)e^{F(r)}dr}} \mbox{    .}
\label{approach:3}
\end{equation}

\item
Examine the behaviour of the Kretschmann scalar \begin{math}R_{\mu\nu\rho\sigma}R^{\mu\nu\rho\sigma}\end{math} at potential singular points.

\end{itemize}

\section{The Weak Field Approximation}

In the weak field, taking \begin{math}B(r)=1+b(r)\end{math} and \begin{math}A(r)=1+a(r)\end{math} with \begin{math}|b(r)| << 1\end{math} and \begin{math}|a(r)|<<1\end{math}, (\ref{approach:1}) can be written as

\begin{equation}
a' \approx \frac{b''r^2 + 2b'r - 2a}{2r} \mbox{    .}
\label{weak:2}
\end{equation}

\noindent This has the solution (where \begin{math}c\end{math} is a constant of integration)
\begin{equation}
a = \frac{b'r}{2} + \frac{c}{r} \mbox{    .}
\label{weak:3}
\end{equation}

\noindent Noting that \begin{math}c=0\end{math} to ensure that \begin{math}A\end{math} is finite at  \begin{math}r=0\end{math}, any choice of \begin{math}B(r)\end{math} for which \begin{math}b'r\end{math} is small for all r would give suitable geon solutions in the weak field.  

For our choice of \begin{math}B(r)\end{math} this becomes

\begin{equation}
a = \frac{mr^5 - 4m^2 l^2 r^2}{(r^3 + 2m l^2)^2} \mbox{    .}
\label{weak:4}
\end{equation}

\begin{figure}[ht]
\includegraphics{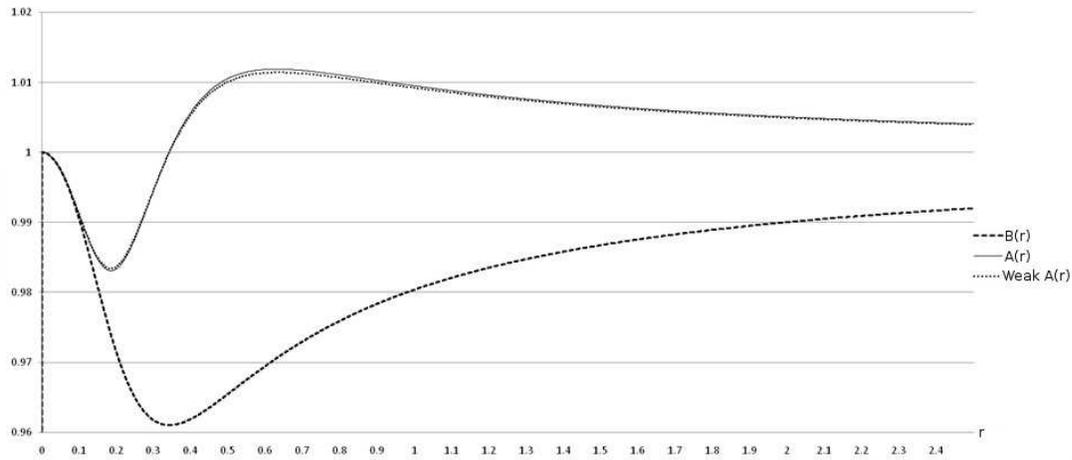}
\caption{\footnotesize B(r) and A(r) plotted against radius r for $l = 1$, $m = 0.01$.  A 4th order numerical iteration of (\ref{approach:1}) is labelled $A(r)$ and the analytical weak field result is labelled $weak A(r)$.  These two approaches give almost identical results for $\frac{m}{l} << 1$.}
\label{fig:weak_A}
\end{figure}

\noindent For \begin{math}r << \left(ml^2 \right)^{\frac{1}{3}}\end{math} we have \begin{math}B \approx 1 + \left( \frac{r}{l} \right)^2\end{math} and \begin{math}A \approx 1 - \left( \frac{r}{l} \right)^2\end{math}.  \begin{math}A(r)\end{math} therefore satisfies the requirements that \begin{math}A(0) = 1\end{math} and \begin{math}\displaystyle\lim_{r \to 0}A(r) = 1 + O(r^n) \mbox{ with } n \ge 2\end{math}.  It is worth noting that for large \begin{math}r\end{math}, \begin{math}A\end{math} differs from Schwarzschild with \begin{math}A \approx 1 + \frac{m}{r}\end{math}.  Furthermore, the conditions for the weak field,

\begin{equation}
|b(r)| = \frac{2mr^2}{r^3+2ml^2} << 1 \mbox{    and}
\label{weak:5}
\end{equation}

\begin{equation}
|a(r)| = \left|\frac{mr^5-4m^2l^2r^2}{(r^3+2ml^2)^2}\right| << 1 
\label{weak:6}
\end{equation}

\noindent are both met for \begin{math}\frac{r}{l} << 1\end{math} and for \begin{math}\frac{m}{l} << 1\end{math}.  The solutions (see fig.\ref{fig:weak_A}) correspond to gravitational geons.

\section{General Behaviour}

We did not find a general solution to (\ref{approach:1}) for strong gravitational fields.  We examined the behaviour of \begin{math}A(r)\end{math} using a fourth order Runge-Kutta numerical iteration.  Using this method, we were unable to begin at exactly \begin{math}A(0) =1\end{math}.  However, \begin{math}A \approx 1 - \left( \frac{r}{l} \right)^2\end{math} for small \begin{math}r\end{math}.  Thus, if we choose an iteration length \begin{math}\Delta r\end{math} so that \begin{math}\frac{\Delta r}{l} = 10^{-5}\end{math} then \begin{math}A(\Delta r) \approx 1\end{math} to within \begin{math}10^{-10}\end{math} (independent of \begin{math}m\end{math}).  Also, having made the observation from direction field plots that the solution is insensitive to small changes in the initial value of A, we chose \begin{math}\Delta r\end{math} appropriately and used \begin{math}A(\Delta r) = 1\end{math} as the starting point of the iteration.

We are interested in solutions for which both \begin{math}A(r)\end{math} and \begin{math}B(r)\end{math} are greater than zero and finite for all \begin{math}r \ge 0\end{math}.  The zeros of \begin{math}B\end{math} can be determined by noting that \begin{math}B(0) = 1\end{math}, \begin{math}\displaystyle\lim_{r \to \infty} B(r) = 1\end{math} and that \begin{math}B' = 0\end{math} at \begin{math}r=(4ml^2)^{\frac{1}{2}}\end{math} where \begin{math}B\end{math} has its minimum value of \begin{math}1-\frac{1}{3}(4\frac{m}{l})^{\frac{2}{3}}\end{math}.  Thus

\begin{eqnarray}
B \mbox{ has } 
  \cases{
   \mbox{no zeros} & if $\frac{m}{l} < \frac{\sqrt{27}}{4}$ \cr
   \mbox{one zero} & if $\frac{m}{l} = \frac{\sqrt{27}}{4}$ \cr
   \mbox{two zeros} & if $\frac{m}{l} > \frac{\sqrt{27}}{4}$  \mbox{.}}
\label{general:10}
\end{eqnarray}

We proceeded to examine the numerical iteration of \begin{math}A(r)\end{math}.  We found that it works well provided we choose \begin{math}m \mbox{ and } l\end{math} for which \begin{math}B'r+4B \ne 0\end{math} for all \begin{math}r > 0\end{math} but that it is unable to navigate instabilities in the solution otherwise.  The zeros of \begin{math}B'r+4B\end{math} are found by observing that \begin{math}B'r+4B \big|_{r=0} = 4 \mbox{, } \displaystyle\lim_{r \to \infty} B'r+4B = 4\end{math} and that \begin{math}\frac{d}{dr}(B'r+4B) = 0\end{math} at \begin{math}r=(2ml^2)^{\frac{1}{3}}\end{math} where \begin{math}B'r+4B\end{math} has its minimum value of \begin{math}4-(\frac{27m}{4l})^{\frac{2}{3}}\end{math}.  Thus

\begin{eqnarray}
B'r+4B \mbox{ has } 
  \cases{
   \mbox{no zeros} & if $\frac{m}{l} < \frac{32}{27}$ \cr
   \mbox{one zero} & if $\frac{m}{l} = \frac{32}{27}$ \cr
   \mbox{two zeros} & if $\frac{m}{l} > \frac{32}{27}$  \mbox{.}}
\label{general:11}
\end{eqnarray}

\noindent Noting that there will be choices of \begin{math}m\end{math} and \begin{math}l\end{math} for which there will be zeros of \begin{math}B'r+4B\end{math} but no zeros of \begin{math}B\end{math}, we investigated the behaviour of \begin{math}A(r)\end{math} around the zeros of \begin{math}B'r+4B\end{math}.

In general, from (\ref{approach:2}), we have

\begin{equation}
f(r)= \frac{2BB''r^2-(B')^2r^2+4BB'r+4B^2}{Br(B'r+4B)} \mbox{    and}
\label{general:2}
\end{equation}

\begin{equation}
g(r)= \frac{-4B}{r(B'r+4B)} \mbox{    .}
\label{general:3}
\end{equation} 

\noindent Taking \begin{math}x = r-r_0\end{math} where \begin{math}B'r+4B \big|_{r=r_0}=0\end{math} we get series approximations for (\ref{general:2}) and (\ref{general:3})

\begin{equation}
f(x) = \frac{\alpha_1}{x} + \alpha_2 + \alpha_3 x +O(x^2)  \mbox{    and}
\label{general:12}
\end{equation} 

\begin{equation}
g(x) = \frac{\beta_1}{x} + \beta_2 + \beta_3 x + O(x^2)  \mbox{    .}
\label{general:13}
\end{equation}  

\noindent In this case, for \begin{math}\alpha_1 \notin \{0, -1, -2, ...\} \end{math}, solutions to (\ref{approach:3}) are of the form

\begin{equation}
\hspace*{-2cm} A(x)= \frac{-1}{\frac{\beta_1}{\alpha_1} - \frac{\beta_1 \alpha_2 - \beta_2 \alpha_1}{\alpha_1(\alpha_1 + 1)} x + \lambda |x|^{-\alpha_1} + O(x^2, \lambda x|x|^{-\alpha_1})} \quad \mbox{(}\lambda \mbox{ - const. of integration). }
\label{general:4}
\end{equation}

\noindent For \begin{math}\alpha_1 \in \{0, -1, -2, ...\} \end{math} we get a different solution for each choice of \begin{math}\alpha_1\end{math}.  For example, choosing \begin{math}\alpha_1 = -1\end{math} gives

\begin{equation}
\hspace*{-2cm} A(x)= \frac{-1}{-\beta_1 + (\beta_2 + \beta_1 \alpha_2) x ln|x| + \beta_1 \alpha_2 x + \lambda x + O(x^2 ln|x|)}  \quad \mbox{(}\lambda \mbox{ - const. of integration). }
\label{general:6}
\end{equation}

\noindent In each case, the dominant term in the denominator as \begin{math}x \to 0\end{math} results in the same outcome as (\ref{general:4}) that

\begin{eqnarray}
\displaystyle\lim_{x \to 0} A(x) = 
  \cases{
    \frac{-\alpha_1}{\beta_1} & if $\alpha_1 \le 0$ or $\lambda = 0$ \cr
    0 & if $\alpha_1 \ge 0$ and $\lambda \neq 0$ .}
\label{general:5}
\end{eqnarray}

\begin{figure}[ht]
\includegraphics{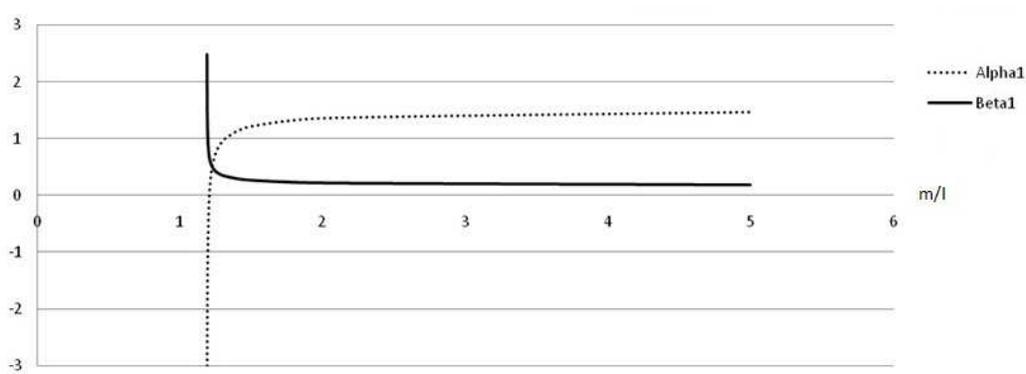}
\caption{\footnotesize $\alpha_1$ and $\beta_1$ plotted against $\frac{m}{l}$ for the first zero of $B'r+4B$.}
\label{fig:AlphasAndBetas}
\end{figure}

\begin{figure}[ht]
\includegraphics{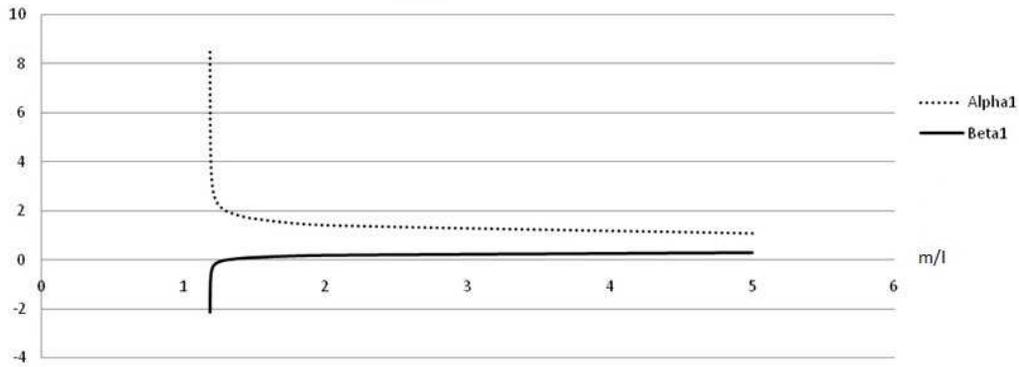}
\caption{\footnotesize $\alpha_1$ and $\beta_1$ plotted against $\frac{m}{l}$ for the second zero of $B'r+4B$.}
\label{fig:AlphasAndBetas2}
\end{figure}

We calculated values for \begin{math}\alpha_1 \mbox{ and } \beta_1\end{math} for each zero of \begin{math}B'r+4B\end{math} given a variety of values of \begin{math}m \mbox{ and } l\end{math} (see figs. \ref{fig:AlphasAndBetas} and \ref{fig:AlphasAndBetas2}).  Having verified our results for a range of choices \begin{math}0.02 < l < 750\end{math}, we found that at each zero, the values of \begin{math}\alpha_1 \mbox{ and } \beta_1\end{math} depend on \begin{math}\frac{m}{l}\end{math} but not on \begin{math}m \mbox{ and } l\end{math} individually.  Knowing that zeros of \begin{math}A(r)\end{math} correspond to singularities in the space-time, we sought to determine where \begin{math}\alpha_1 = 0\end{math} in terms of \begin{math}\frac{m}{l}\end{math}.

With \begin{math}x=r-r_0\end{math}, we have \begin{math}B'r+4B \approx x(B''r+5B')\big|_{r=r_0}\end{math} in the vicinity of each zero of \begin{math}B'r+4B\end{math} so  

\begin{eqnarray}
\alpha_1&= \frac{2B''r-\frac{(B')^2r}{B}+4B'+\frac{4B}{r}}{B''r+5B'} \bigg|_{r=r_0} \nonumber \\
&=\frac{2B''r+7B'}{B''r+5B'} \bigg|_{r=r_0} \mbox{    .}
\label{general:7}
\end{eqnarray}

\noindent For \begin{math}\alpha_1 = 0\end{math}, substituting in (\ref{introduction:13}), we get the following quadratic in \begin{math}r^3\end{math} 

\begin{equation}
r^6 + 14ml^2r^3-24m^2l^4 = 0
\label{general:1}
\end{equation}

\noindent with one positive real root corresponding to \begin{math}r \approx (1.544ml^2)^{\frac{1}{3}}\end{math}.  Substituting this into \begin{math}B'r+4B=0\end{math} yields \begin{math}\frac{m}{l} \approx 1.202\end{math}.  Based on this (and again looking at figs. \ref{fig:AlphasAndBetas} and \ref{fig:AlphasAndBetas2}), we know that for solutions with \begin{math}\frac{m}{l}>\frac{32}{27}\end{math}, there will be two zeros of \begin{math}B'r+4B\end{math}, that the first will have \begin{math}\alpha_1 > 0\end{math} for  \begin{math}\frac{m}{l} > 1.202\end{math} and the second will always have \begin{math}\alpha_1 > 0\end{math}.  

We know that \begin{math}A' \to \pm \infty\end{math} corresponds to singular behaviour in the space-time.  Differentiating (\ref{general:4}) gives

\begin{equation}
A'(x)= \frac{ - \frac{\beta_1 \alpha_2 - \beta_2 \alpha_1}{\alpha_1(\alpha_1 + 1)} - \frac{\alpha_1 \lambda |x|^{-\alpha_1}}{x} + ... }{\left(\frac{\beta_1}{\alpha_1} - \frac{\beta_1 \alpha_2 - \beta_2 \alpha_1}{\alpha_1(\alpha_1 + 1)}x + \lambda |x|^{-\alpha_1} + ...\right)^2}  
\label{general:8}
\end{equation}

\begin{eqnarray}
\displaystyle\lim_{x \to 0} A'(x) =
  \cases{
   \frac{\alpha_1(\beta_2 \alpha_1 - \beta_1 \alpha_2)}{\beta_1^2(\alpha_1 + 1)}  & if $\alpha_1 < -1$ or $\lambda = 0$ \cr
   -sgn(\alpha_1 \lambda x) \infty  & if $-1 < \alpha_1 < 1$  and $\lambda \neq 0$ \cr
   0  & if $\alpha_1 > 1$ and $\lambda \neq 0$ .}
\label{general:9}
\end{eqnarray}

We know from (\ref{general:5}) that we need not consider \begin{math}\alpha_1 > 0\end{math} (unless \begin{math}\lambda = 0\end{math}), but the behaviour of \begin{math}A(r)\end{math} does depend on whether \begin{math}\alpha_1\end{math} is greater or less than -1 at the zeros $B'r+4B$.  We therefore sought to determine where  \begin{math}\alpha_1 = -1\end{math} in terms of \begin{math}\frac{m}{l}\end{math} and numerically found this corresponds to \begin{math}\frac{m}{l} \approx 1.191\end{math}.

Now we found that at each zero, the values of \begin{math}\alpha_1 \mbox{ and } \beta_1\end{math} depend on \begin{math}\frac{m}{l}\end{math} but not on \begin{math}m \mbox{ and } l\end{math} individually.  That is not true of the values of \begin{math}\alpha_2 \mbox{ and } \beta_2\end{math}.  However, we did determine (again verifying our results for a range of choices \begin{math}0.02 < l < 750\end{math}) that \begin{math}sgn(\alpha_2)\end{math}, \begin{math}sgn(\beta_2)\end{math} and the ratio \begin{math}\frac{\alpha_2}{\beta_2}\end{math} at each zero depends on \begin{math}\frac{m}{l}\end{math} but not on \begin{math}m \mbox{ and } l\end{math} individually.  We can therefore say that the sign of \begin{math}\frac{\alpha_1(\beta_2 \alpha_1 - \beta_1 \alpha_2)}{\beta_1^2(\alpha_1 + 1)} = \frac{\alpha_1 \beta_2 \left( \alpha_1 - \beta_1 \frac{\alpha_2}{\beta_2} \right) }{\beta_1^2(\alpha_1 + 1)}\end{math} depends on \begin{math}\frac{m}{l}\end{math} but not on \begin{math}m \mbox{ and } l\end{math} individually.

Having determined how the choice of \begin{math}\frac{m}{l}\end{math} affects the behaviour of \begin{math}A\end{math} and \begin{math}A'\end{math} at each zero of \begin{math}B'r+4B\end{math}, we looked at each distinct range of \begin{math}\frac{m}{l}\end{math} values considering both \begin{math}\lambda = 0\end{math} and \begin{math}\lambda \ne 0\end{math} solutions.  For each case, we sought to determine if our solution corresponds to gravitational geons.   

\subsection{\texorpdfstring{$\frac{m}{l} < \frac{32}{27}$ -- Solutions correspond to gravitational geons}{m/l < 1.185}}

\begin{figure}[ht]
\includegraphics{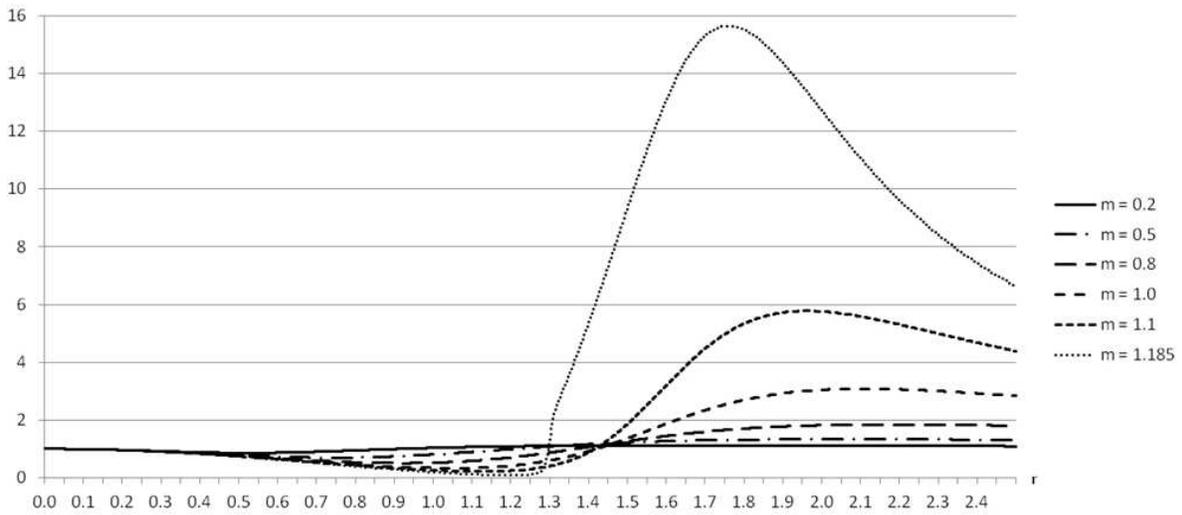}
\caption{\footnotesize Function A(r) plotted against radius r.  Here we choose $l = 1$ and chart a variety of values for \textit{m} in the specified region.  A 4th order Runge-Kutta numerical iteration is used to generate each result.}
\label{fig:A_no_critical_point}
\end{figure}

\begin{figure}[ht]
\includegraphics{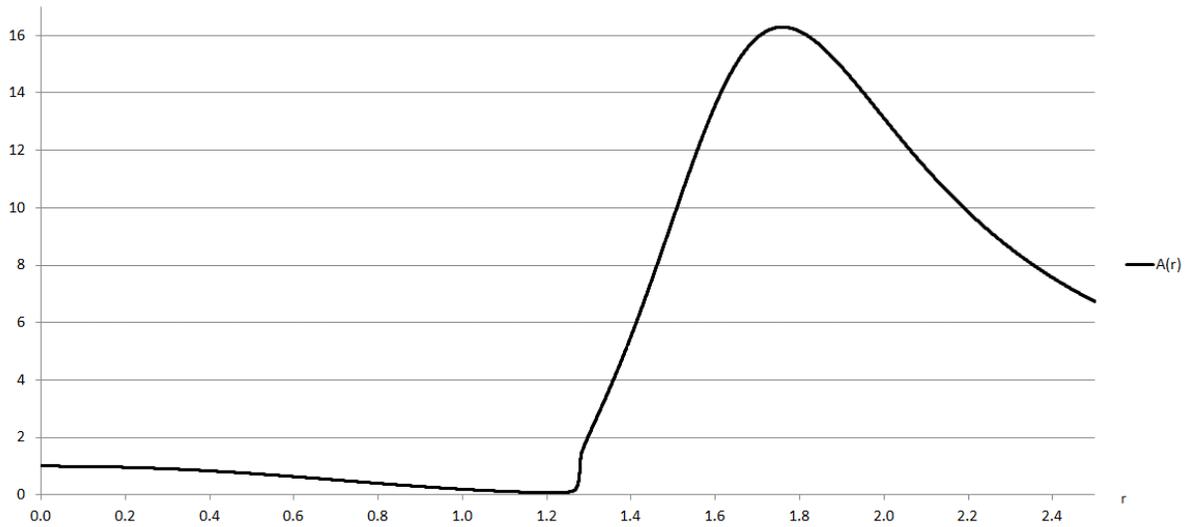}
\caption{\footnotesize The function A(r) plotted against radius r for $l = 1$, $m = 1.189$.  A 4th order Runge-Kutta numerical iteration was unable to navigate the zeros of $B'r + 4B$ (at  $r = 1.281$, $1.392$), so the result is constructed piecewise between these points.}
\label{fig:A_1.189}
\end{figure}

For these values of \begin{math}\frac{m}{l}\end{math}, $B'r+4B$ has no zeros.  Consequently, all solutions with \begin{math} A(0) = 1 \end{math} correspond to gravitational geons.  Choosing \begin{math} l = 1 \end{math} we confirmed the behavior of \begin{math} A \end{math} using iterative numerical analysis (see fig.\ref{fig:A_no_critical_point}).  Also evident is the progression from the weak field solution for \begin{math}\frac{m}{l} << 1\end{math} (fig.\ref{fig:weak_A}) to the piecewise constructed solution for \begin{math}\frac{32}{27} < \frac{m}{l} < 1.191\end{math} (fig.\ref{fig:A_1.189}) which we discuss in the following section.
  
\subsection{\texorpdfstring{$\frac{32}{27} < \frac{m}{l} < 1.191$ with $\lambda = 0$ at the second zero of $B'r+4B$}{1.185 < m/l < 1.191}}

At the first zero of $B'r+4B$, \begin{math}A = \frac{-\alpha_1}{\beta_1} > 0\end{math} and \begin{math}A'=\frac{\alpha_1(\beta_2 \alpha_1 - \beta_1 \alpha_2)}{\beta_1^2(\alpha_1 + 1)} > 0\end{math} for values of $\frac{m}{l}$ in this range.  There is therefore no singular behaviour and no horizon indicated at the corresponding spherical hypersurface.  

At the second zero of $B'r+4B$, \begin{math}A=0\end{math} indicating singular behaviour in the space-time except when $\lambda = 0$.  In that case, \begin{math}A = \frac{-\alpha_1}{\beta_1} > 0\end{math} and \begin{math}A'=\frac{\alpha_1(\beta_2 \alpha_1 - \beta_1 \alpha_2)}{\beta_1^2(\alpha_1 + 1)} > 0\end{math} at the second zero.  Consequently, no singular behaviour or horizon is indicated.

\begin{figure}[ht]
\includegraphics{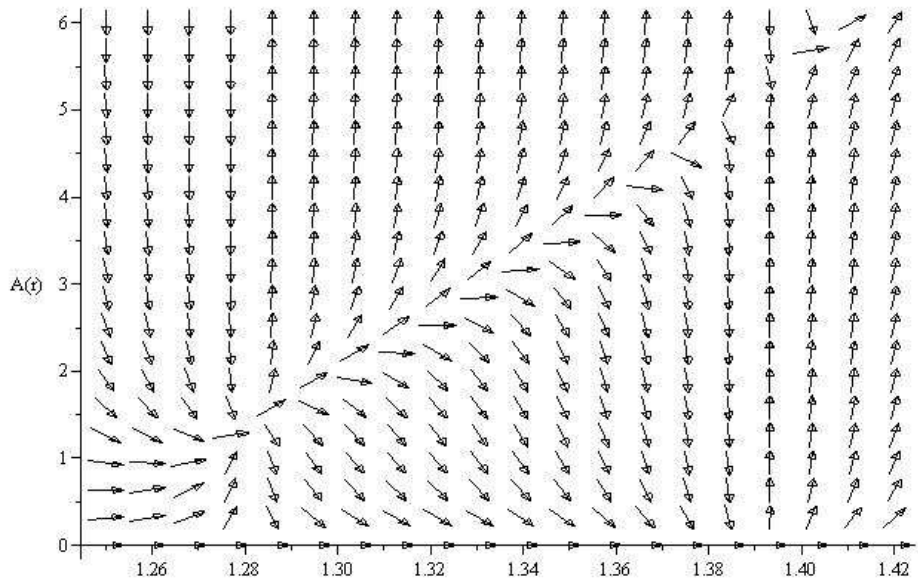}
\caption{\footnotesize This direction field plot of A(r) against radius r shows that A(r) is numerically unstable at the zeros of $B'r + 4B$.  Here we choose $l = 1$, $m = 1.189$ such that the zeros are found at  $r = 1.281$, $1.392$.}
\label{fig:A_direction_plot}
\end{figure}

\begin{figure}[ht]
\includegraphics{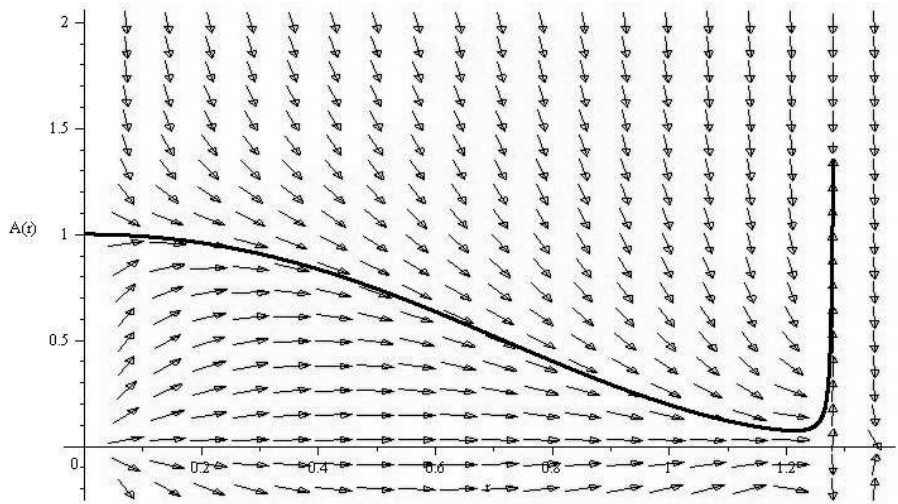}
\caption{\footnotesize Forward iteration of A(r) plotted from $r = 0$ to the first zero of $B'r + 4B$ for $l = 1$, $m = 1.189$.  $A \to \frac{-\alpha_1}{\beta_1} \approx 1.38$ as the solution approaches the first zero.}
\label{fig:A_1.189_FirstForward}
\end{figure}

\begin{figure}[ht]
\includegraphics{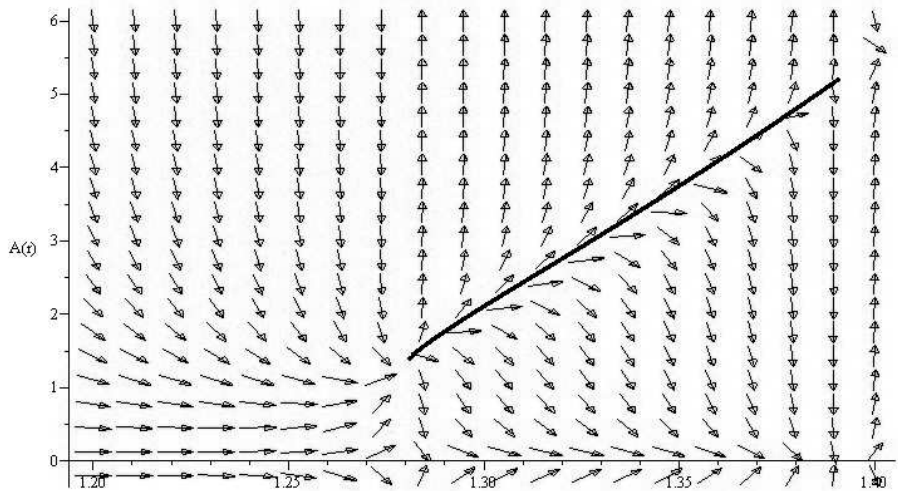}
\caption{\footnotesize Reverse iteration of A(r) plotted from the second zero of $B'r + 4B$ to the first for $l = 1$, $m = 1.189$.}
\label{fig:A_1.189_SecondBackward}
\end{figure}

\begin{figure}[ht]
\includegraphics{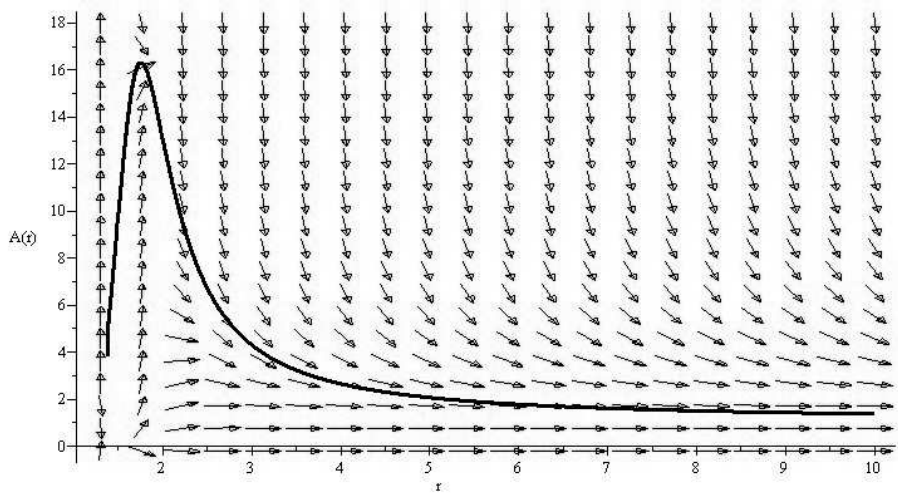}
\caption{\footnotesize Forward iteration of A(r) plotted from the second zero of $B'r + 4B$ for $l = 1$, $m = 1.189$.}
\label{fig:A_1.189_SecondForward}
\end{figure}

Whereas our analysis of the Kretshmann scalar shows the possibility of solutions corresponding to gravitational geons with \begin{math}\lambda = 0\end{math} at the second zero, instability at each zero prevents us from confirming any particular solution using iterative methods (see fig.\ref{fig:A_direction_plot}).  However, we have been able to confirm that \begin{math}\displaystyle\lim_{x \to 0} A = \frac{-\alpha_1}{\beta_1}\end{math} and \begin{math}\displaystyle\lim_{x \to 0} A' = \frac{\alpha_1(\beta_2 \alpha_1 - \beta_1 \alpha_2)}{\beta_1^2(\alpha_1 + 1)}\end{math} around each zero for each section constructed numerically (see figs. \ref{fig:A_1.189_FirstForward}, \ref{fig:A_1.189_SecondBackward}, \ref{fig:A_1.189_SecondForward}).

We piecewise constructed \begin{math}A\end{math} around each zero of \begin{math}B'r+4B\end{math} using the numerically constructed sections (see fig.\ref{fig:A_1.189}).  The resulting solution is \emph{sewn together} at the spherical hypersurfaces corresponding to each zero.

The sewing together of two manifolds may induce a surface energy-momentum tensor \cite{Israel:1966rt,Misner:1973}

\begin{equation}
S_{\mu\nu} = \frac{1}{8 \pi} \left( [K_{\mu\nu}] - [K] h_{\mu\nu} \right)
\label{general:15}
\end{equation}

\noindent on the surface where they are joined.  Here \begin{math}K_{\mu\nu}\end{math} is the extrinsic curvature of the surface, \begin{math}K= K^{\mu}{}_{\mu}\end{math}, \begin{math}h_{\mu\nu}\end{math} is the induced metric on the surface and \begin{math}[K_{\mu\nu}]\end{math} denotes the jump in \begin{math}K_{\mu\nu}\end{math} across the surface.

Now, \begin{math}K_{\mu\nu}\end{math} depends on the metric and its first derivatives \cite{Carroll:2004}.  Given that \begin{math}A \mbox{, } B \mbox{ and } B'\end{math} are all continuous in our piecewise constructed solution, there will be a non-vanishing \begin{math}S_{\mu\nu}\end{math} only if \begin{math}A'\end{math} is discontinuous across either hypersuface corresponding to a zero of \begin{math}B'r + 4B\end{math}.  Our analysis shows that \begin{math}A'\end{math} is continuous implying that \begin{math}S_{\mu\nu} = 0\end{math}.  

It should be noted that there could be contributions to \begin{math}S_{\mu\nu}\end{math} from the Weyl term (\ref{introduction:14}).  If that were the case, \begin{math}S_{\mu\nu} = 0\end{math} in our piecewise constructed solution would imply that there is a localized gravitational source (such as matter) at the hypersurface and the stresses contributed by each cancel one another exactly.  Since the space-time off the brane is not known, it is not possible to check to see if such contributions exist.

If there are no contributions to \begin{math}S_{\mu\nu}\end{math} from the Weyl tensor, the piecewise constructed solutions correspond to gravitational geons.

\subsection{\texorpdfstring{Choices of $\frac{m}{l}$ for which no gravitational geon solutions exist}{m/l > 1.191}}

Our analysis shows that the Kretshmann scalar diverges at the spherical hypersurfaces corresponding to the zeros of $B'r+4B$ for values of $\frac{m}{l} > 1.202$.  We have therefore determined that solutions that correspond to gravitational geons do not exist for these values of $\frac{m}{l}$.  

In the range $1.191 < \frac{m}{l} < 1.202$, divergence of the Kretshmann scalar rules out the possibility of gravitational geon solutions, except in the special case that $\lambda = 0$ at both zeros of $B'r+4B$. However, direction field plots indicate that these solutions are inconsistent with our boundary conditions.

As discussed in the previous section, gravitational geon solutions are again ruled out by divergence of the Kretshmann scalar in the range $\frac{32}{27} < \frac{m}{l} < 1.191$, except in the special case that $\lambda = 0$ at the second zero of $B'r+4B$. 

There are therefore no solutions that correspond to gravitational geons possible for $\frac{m}{l} > \frac{32}{27}$ with the exception of $\lambda = 0$ at the second zero of $B'r+4B$ for $\frac{32}{27} < \frac{m}{l} < 1.191$. 

\section{Conclusion}

In this paper, we investigated a particular static, spherically symmetric 3+1 dimensional space-time for behaviour consistent with that of gravitational geons on a brane embedded in a 4+1 dimensional space-time.  In the absence of a general solution, we investigated the behaviour of the space-time using a variety of methods and were able to ascertain that, for a particular set of parameter choices, gravitational geons are possible.  

The significance of the result presented in this paper is that, in general, for brane-world space-times gravitational geons are possible.  We investigated solutions in the weak field limit and showed that gravitational geons will exist as long as \begin{math}B'r\end{math} is small for all \begin{math}r \ge 0\end{math}.  We also investigated the specific function \begin{math}B(r)=1-\frac{2mr^2}{r^3+2ml^2} \end{math} and showed that gravitational geon solutions exist for all parameter choices such that \begin{math}\frac{m}{l} < \frac{32}{27}\end{math}.  These solutions include both weak field and strong field geons.

If our universe corresponds to the Randall-Sundrum model, then gravitational geons could represent a form of dark matter (this possibility has also been discussed by Sones \cite{Sones:2005gn} for quantum geons with a Klein-Gordon field).

\ack

This research was supported by the Natural Sciences and Engineering Research Council of Canada.

\newpage

\section*{Figures}

\end{document}